\documentstyle[11pt,newpasp,twoside,epsf]{article}
\def\edcomment#1{\iffalse\marginpar{\raggedright\sl#1\/}\else\relax\fi}
\reversemarginpar

\begin{document}
\title{The EMBH model in GRB~991216 and GRB~980425}
\author{R. Ruffini, C.L. Bianco, P. Chardonnet, F. Fraschetti, S.-S. Xue}
\affil{ICRA --- International Center for Relativistic Astrophysics and Dipartimento di Fisica, Universit\`a di Roma ``La Sapienza'', Piazzale Aldo Moro 5, I-00185 Roma, Italy; ruffini@icra.it, bianco@icra.it, chardon@lapp.in2p3.fr, fraschetti@icra.it, vitagliano@icra.it, xue@icra.it}

\begin{abstract}
This is a summary of the two talks presented at the Rome GRB meeting by C.L. Bianco and R. Ruffini. It is shown that by respecting the Relative Space-Time Transformation (RSTT) paradigm and the Interpretation of the Burst Structure (IBS) paradigm, important inferences are possible: a) in the new physics occurring in the energy sources of GRBs, b) on the structure of the bursts and c) on the composition of the interstellar matter surrounding the source.
\end{abstract}

The understanding of new astrophysical phenomena is reached as soon as its energy source has been identified. This has been the case for pulsars (see Hewish et al., 1968) where the rotational energy of the neutron star was identified as the energy source (see e.g. Gold, 1968, 1969). Similarly, in binary X-ray sources the accretion process from a normal companion star in the deep potential well of a neutron star or a black hole has clearly pointed to the gravitational energy of the accreting matter as the basic energy source and consequently all the main features of the light curves of the sources have been clearly understood (Giacconi \& Ruffini, 1978). In this spirit, our work in the field of Gamma-Ray Bursts (GRBs) has focused on identifing the energy extraction process from the black hole (Christodoulou \& Ruffini, 1971) as the basic energy source for the GRB phenomenon. The distinguishing feature of this approach is a theoretically predicted source energetics all the way up to $1.8\times 10^{54}\left(M_{BH}/M_{\odot}\right) {\rm ergs}$ for $3.2 M_{\odot} \le M_{BH} \le 7.2 \times 10^6 M_{\odot}$ (Damour \& Ruffini, 1975). In particular, the formation of a ``dyadosphere'', during the gravitational collapse leading to a black hole endowed with electromagnetic structure (EMBH) has been indicated as the initial boundary conditions of the GRB process (Ruffini, 1998; Preparata et al., 1998). Our model has been referred as ``the EMBH model for GRBs'', although the EMBH physics only determines the initial boundary conditions of the GRB process by specifying the physical parameters and spatial extension of the neutral electron positron plasma originating the phenomenon.

Traditionally, following the observations of the {\em Vela} (Strong, 1975) and {\em CGRO}\footnote{see http://cossc.gsfc.nasa.gov/batse/} satellites, GRBs have been characterized by few parameters such as the fluence, the characteristic duration ($T_{90}$ or $T_{50}$) and the global time averaged spectral distribution (Band et al., 1993). With the observations of {\em BeppoSAX}\footnote{see http://www.asdc.asi.it/bepposax/} and the discovery of the afterglow, and the consequent optical identification, the distance of the GRB source has been determined and consequently the total energetics of the source has been added as a crucial parameter.

The observed energetics of GRBs computed for spherically symmetric explosions do coincide with the ones theoretically predicted in Damour \& Ruffini (1975). This fact has convinced us of the necessity to develop in full details the EMBH model. For simplicity, we have considered the vacuum polarization process occurring in an already formed Riessner-Nordstr\"om black hole (Ruffini, 1998; Preparata et al., 1998), whose dyadosphere has an energy $E_{dya}$. It is clear, however, that this is only an approximation to the real dynamical description of the process of gravitational collapse to an EMBH. In order to prepare the background for attacking this extremely complex dynamical process, we have clarified some basic theoretical issues, necessary to implement the description of the fully dynamical process of gravitational collapse to an EMBH (see Ruffini \& Vitagliano, 2002, 2003; Cherubini et al., 2002). We have then given the constitutive equations for the five eras in our model. {\em The Era I}: the $e^+e^-$ pairs plasma, initially at $\gamma=1$, self propels itself away from the dyadosphere as a sharp pulse (the PEM pulse), reaching Lorentz gamma factor of the order of 100 (Ruffini et al., 1999). {\em The Era II}: the PEM pulse, still optically thick, engulfs the remnant left over in the process of gravitational collapse of the progenitor star with a drastic reduction of the gamma factor; the mass $M_B$ of this engulfed baryonic material is expressed by the dimensionless parameter $B=M_Bc^2/E_{dya}$ (Ruffini et al., 2000). {\em The Era III}: the newly formed pair-electromagnetic-baryonic (PEMB) pulse, composed of $e^+e^-$ pair and of the electrons and baryons of the engulfed material, self-propels itself outward reaching in some sources Lorentz gamma factors of $10^3$--$10^4$; this era stops when the transparency condition is reached and the emission of the proper-GRB (P-GRB) occurs (Bianco et al., 2001). {\em The Era IV}: the resulting accelerated baryonic matter (ABM) pulse, ballistically expanding after the transparency condition has been reached, collides at ultrarelativistic velocities with the baryons and electrons of the interstellar matter (ISM) which is assumed to have a average constant number density, giving origin to the afterglow. {\em The Era V}: this era represents the transition from the ultrarelativistic regime to the relativistic and then to the non relativistic ones (Ruffini et al., 2003a).

Our approach differs in many respect from the ones in the current literature. The major difference consists in the appropriate theoretical description of all the above five eras, as well as in the evaluation of the process of vacuum polarization originating the dyadosphere. The dynamical equations as well as the description of the phenomenon in the laboratory time and the time sequence carried by light signals recorded at the detector have been explicitly integrated (see e.g. Ruffini et al., 2003a, 2003b). In doing so we have also corrected a basic conceptual inadequacy, common to all the current works on GRBs, which led to an improper spacetime parametrization of the GRB phenomenon, preempting all these works from their predictive power: the relation between the photon arrival time at the detector and their emission time in the laboratory frame, expressed in our approach by an integral of a function of the Lorentz gamma factor extended over all the past source worldlines, has been in the current literature expressed as a function of an instantaneous value of the Lorentz gamma factor. This two approaches are conceptually very different and lead to significant qualtitative differences (Ruffini et al., 2003a, 2003b). The description of the inner engine originating the GRBs described has never been addressed in the necessary details in the literature. Only the treatment of the afterglow has been widely considered in the literature by the so-called ``fireball model'' (see e.g. M\'esz\'aros \& Rees, 1992, 1993; Rees \& M\'esz\'aros, 1994; Piran, 1999 and references therein).

However, also in the description of the afterglow, there are major differences between the works in the literature and our approach:\\
{\bf a)} Processes of synchrotron radiation and inverse Compton as well as an adiabatic expansion in the source generating the afterglow are usually adopted in the current literature. On the contrary, in our approach: 1) a ``fully radiative'' condition is systematically adopted in the description of the X-ray and $\gamma$-ray emission of the afterglow; 2) the basic microphysical emission process is traced back to the physics of shock waves as considered by Zel'dovich \& Rayzer (1966); and 3) a special attention is given to identify such processes in the comoving frame of the shock front generating the observed spectra of the afterglow (see Ruffini et al., 2003c).\\
{\bf b)} In the current literature the variation of the gamma Lorentz factor during the afterglow is expressed by a unique power-law of the radial co-ordinate of the source and a similar power-law relation is assumed also between the radial coordinate of the source and the asymptotic observer frame time. Such simple approximations appear to be quite inadequate and do contrast with the almost hundred pages summarizing the needed computations which we have recently presented in four long articles (Ruffini et al., 2002a, 2003a, 2003b, 2003d). In our approach the dynamical equations of the source are integrated self-consistently with the constitutive equations relating the observer frame time to the laboratory time and the boundary conditions are adopted and uniquely determined by each previous era of the GRB source (see e.g. Ruffini et al., 2002b, 2003a, 2003b, 2003c).\\
{\bf c)} At variance with the many power-laws for the observed afterglow flux found in the literature, our treatment naturally leads to a ``golden value'' for the bolometric luminosity power-law index $n=-1.6$.

The fit of the EMBH model to the observed afterglow data fixes the only two free parameters of our theory: the $E_{dya}$ and the $B$ parameter, measuring the remnant mass left over by the gravitational collapse of the progenitor star (Ruffini et al., 2002b, 2003a, 2003b, 2003c).

It is not surprising that such large differences in the theoretical treatment have led to a different interpretation of the GRB phenomenon as well as to the identification of new fundamental physical regimes. The introduction of new interpretative paradigms has been necessary and the theory has been confirmed by the observation to extremely high accuracy. 

In particular from the definition of the complete space-time coordinates of the GRB phenomenon as a function of the radial coordinate, the comoving time, the laboratory time, the arrival time and the arrival time at the detector, expressed in Tab.~1 of Ruffini et al. (2002a, 2003a, 2003d), it has been concluded that in no way a description of a given era is possible in the GRB phenomena without the knowledge of the previous ones. Therefore the afterglow as such cannot be interpreted unless all the previous eras have been correctly computed and estimated. It has also become clear that a great accuracy in the analysis of each era is necessary in order to identify the theoretically predicted features with the observed ones. If this is done, the GRB phenomena presents an extraordinary precise correspondence between the theoretically predicted features and the observations leading to the exploration of totally new physical and astrophysical process with unprecedented accuracy. This has been expressed in the relative space-time transformation (RSTT) paradigm: ``the necessary condition in order to interpret the GRB data, given in terms of the arrival time at the detector, is the knowledge of the {\em entire} worldline of the source from the gravitational collapse. In order to meet this condition, given a proper theoretical description and the correct constitutive equations, it is sufficient to know the energy of the dyadosphere and the mass of the remnant of the progenitor star'' (Ruffini et al., 2001a).

Having determined the two independent parameters of the EMBH model, namely $E_{dya}$ and $B$, by the fit of the afterglow we have introduced a new paradigm for the Interpretation of the Burst Structure: the IBS paradigm (Ruffini et al., 2001b). In it we reconsider the relative roles of the afterglow and the burst in the GRBs by defining in this complex phenomenon two new phases:\\
{\bf 1)} the {\em injector phase} starting with the process of gravitational collapse, encompassing the above Eras I, II, III and ending with the emission of the Proper-GRB (P-GRB);\\
{\bf 2)} the {\em beam-target phase} encompassing the above Eras IV and V giving rise to the afterglow. In particular in the afterglow three different regimes are present for the average bolometric intensity : one increasing with arrival time, a second one with an Extended Afterglow Peak Emission (E-APE) and finally one decreasing as a function of the arrival time. Only this last one appears to have been considered in the current literature (Ruffini et al., 2001b).

\begin{figure}
\plottwo{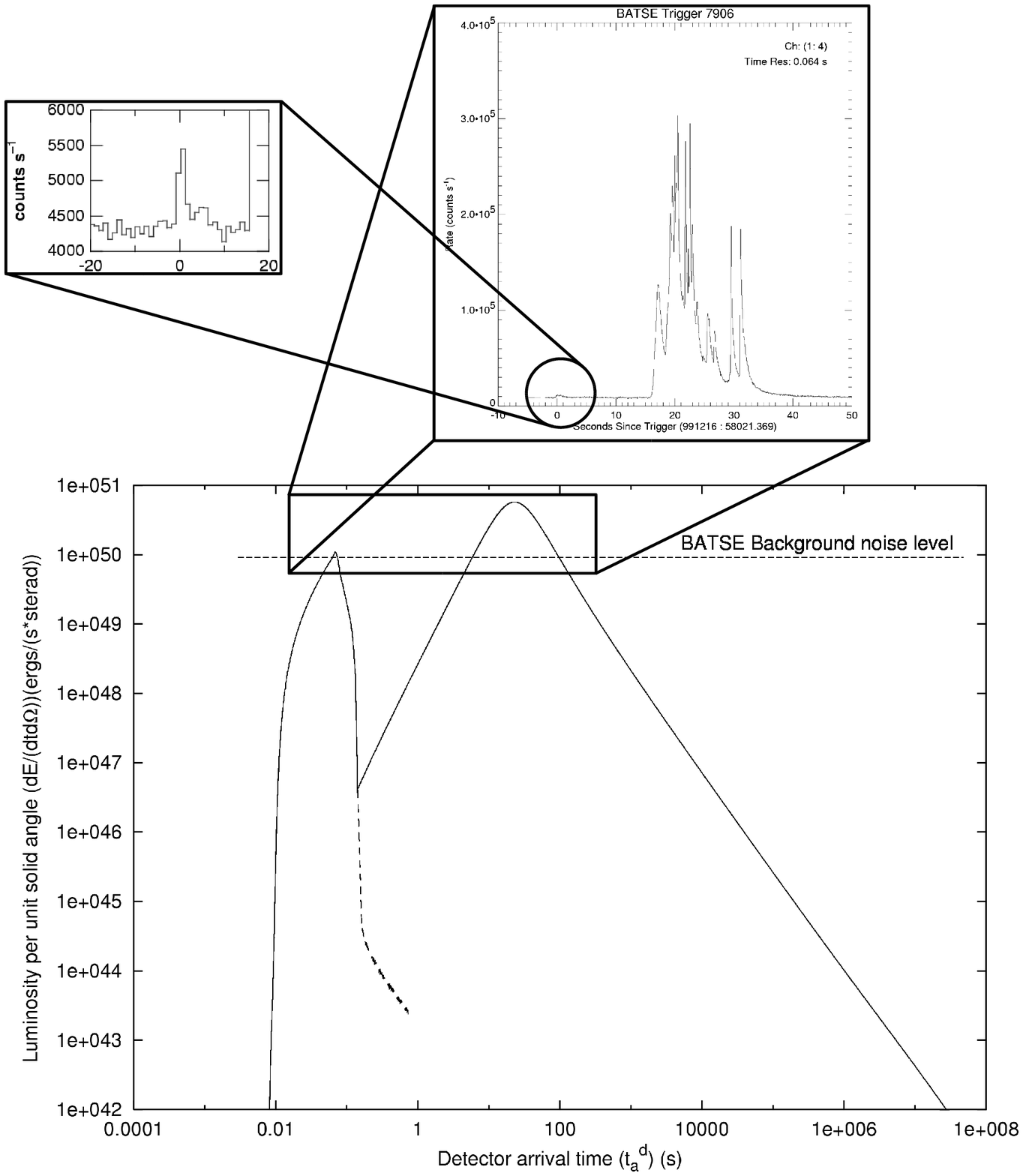}{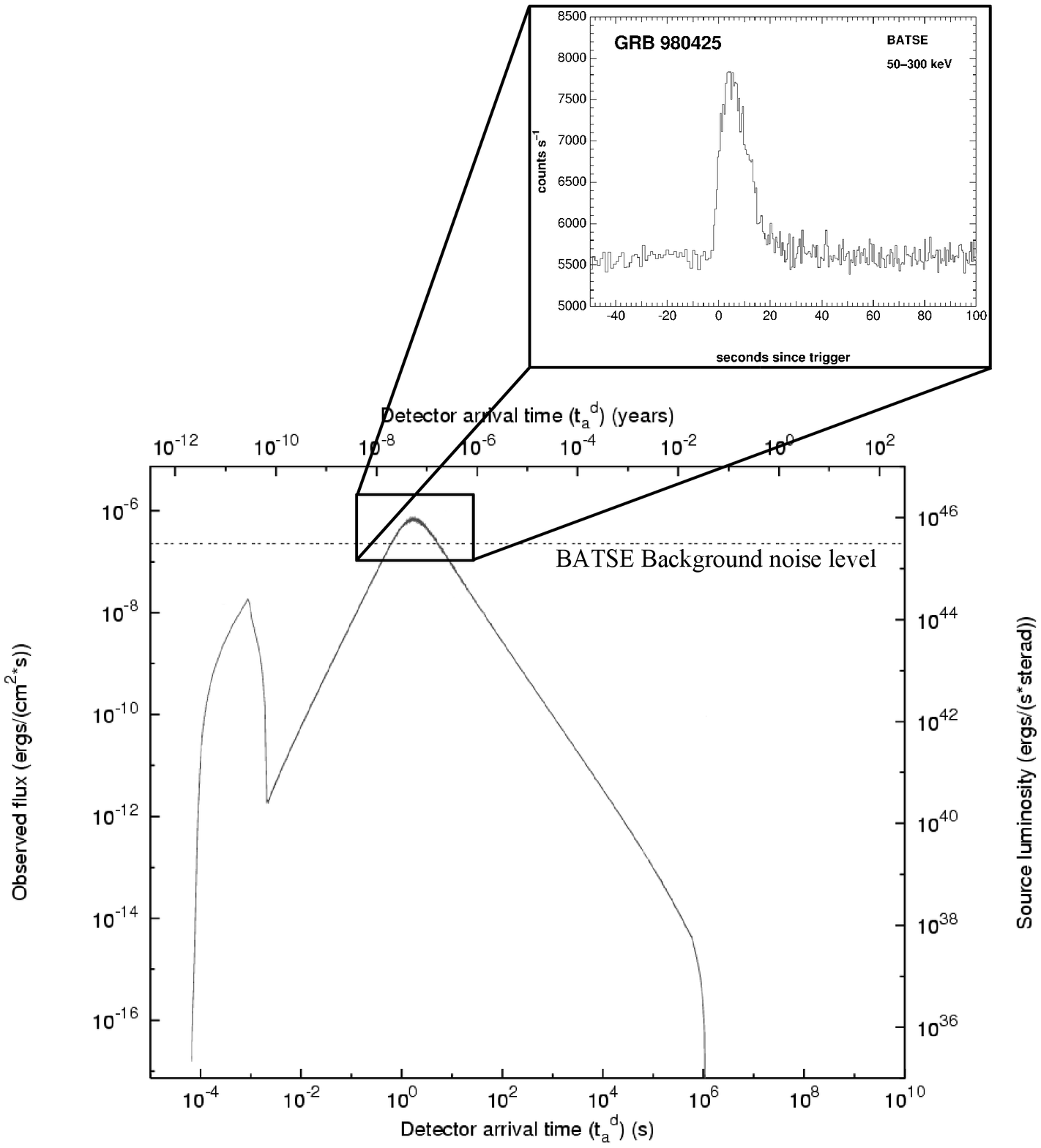}
\caption{The EMBH theory applied to GRB~991216 (left) and GRB~980425 (right). Note the structure of the P-GRB and the E-APE with respect to the BATSE noise threshold.}
\end{figure}

We have applied the EMBH model to GRB~991216 as a prototype. We think that this source will play for GRBs a role similar to the one of NP0532 (the Crab pulsar) in the understanding of the pulsar phenomenon. The GRB~991216 presents clearly the three fundamental aspects: the P-GRB, the E-APE and the late phases of the afterglow (Ruffini et al., 2001a,b). What makes this system so unique are the outstanding data obtained by BATSE in the P-GRB and in the E-APE and concurrently the very high quality ones on the afterglow obtained by R-XTE and Chandra. It is the simultaneous occurrence of these three features which makes this source so attractive and becoming the prototype for GRBs. We have computed the intensity ratio of the afterglow to the P-GRB ($1.45 \cdot 10^{-2}$), and the arrival time of the P-GRB ($8.413 \cdot 10^{-2}$s) as well as the arrival time of the peak of the afterglow ($19.87$s) (see Fig.~1 and Ruffini et al., 2001b, 2002a, 2003a, 2003d). The fact that the theoretically predicted intensities coincide within a few percent with the observed ones and that the arrival time of the P-GRB and the peak of the afterglow also do coincide within a tenth of millisecond with the observed one can be certainly considered a clear success of the predictive power of the EMBH model.

As a by-product of this successful analysis, we have reached the following conclusions:\\
{\bf a)} The most general GRB is composed by a P-GRB, an E-APE and the rest of the afterglow. The ratio between the P-GRB and the E-APE intensities is a function of the B parameter.\\
{\bf b)} In the limit B=0 all the energy is emitted in the P-GRB. These events represent the ``short burst'' class, for which no afterglows has been observed.\\
{\bf c)} The ``long bursts'' do not exist, they are just part of the afterglow, the E-APEs.

Our model has been also applied to GRB~980425 showing not only the agreement with the observed luminosity but also, in particular, that in both sources there is not significant departure from spherical symmetry (see Fig.~1).

While the aboce analysis of the average bolometric luminosity of GRB was performed in the radial approximation, we have also developed the full non-radial approximation, taking into account all the relativistic corrections for the off-axis emission from the spherically symmetric expansion of the ABM pulse (Ruffini et al., 2002b, 2003b). We have so defined the temporal evolution of the ABM pulse visible area, as well as the equitemporal surfaces (see Fig.~2 and Ruffini et al., 2002b, 2003b).

\begin{figure}
\plotone{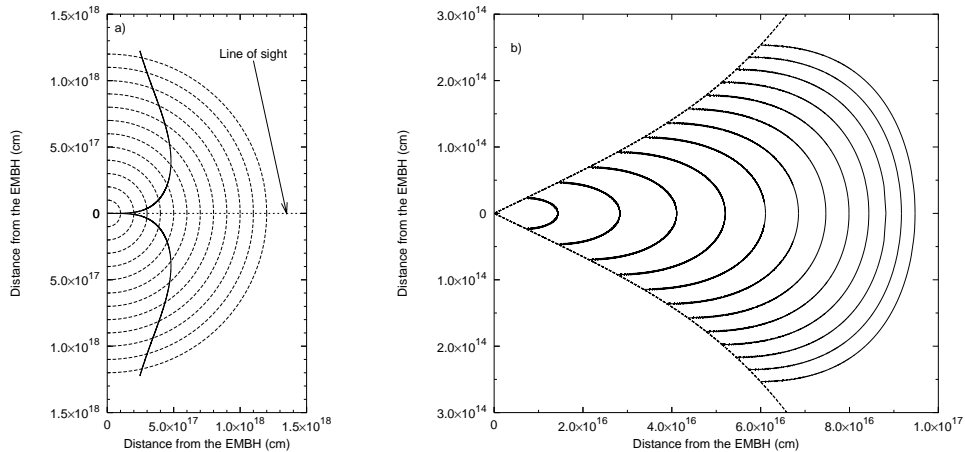}
\caption{The temporal evolution of the visible area of the ABM pulse external surface (left) and the equitemporal surfaces for selected values of the detector arrival time encompassing the E-APE (right) for the case of GRB~991216. Details in Ruffini et al. (2002b, 2003b, 2003d).}
\end{figure}

Having satisfactorily identified the average intensity distribution of the afterglow and the relative position of the P-GRB, in Ruffini et al. (2001c) we have addressed the issue whether the fast temporal variation observed in the so-called long bursts, on time scales as short as fraction of a second (see e.g. Fishman \& Meegan, 1995), can indeed be explained as an effect of inhomogeneities in the interstellar medium. Such a possibility was pioneered in the work by Dermer \& Mitman (1999), purporting that such a time variability corresponds to a tomographic analysis of the ISM. In order to probe the validity of such an explanation, we have first considered the simplified case of the radial approximation (Ruffini et al., 2001c). The aim has been to explain the observed fluctuation in intensity on a fraction of a second as originated from inhomogeneities in ISM, typically of the order of $10^{16}$ due to apparent ``superluminal'' behavior of roughly $10^5$c. These ``superluminal'' effects can be derived consistently from the dynamics of the source. After this successful attempt, we have proceeded to the non-radial approximation, taking into account all the relativistic corrections for the off-axis emission from the spherically symmetric expansion of the ABM pulse (Ruffini et al. 2002b, 2003b). We have so defined the temporal evolution of the ABM pulse visible area, as well as the equitemporal surfaces (EQTS, Ruffini et al. 2002b, 2003b). We have then described the inhomogeneities of the ISM by an appropriate density profile (mask) of an ISM cloud. Of course at this stage,  for simplicity, only the case of spherically symmetric ``spikes'' with over-density separated by low-energy regions, has been considered. Each spike has been assumed to have the spatial extension of $10^{15}$cm. The cloud average density is $<n_{ism}> = 1\, {\rm particle}/{\rm cm}^3$. The corresponding analysis for GRB~980425 has been presented in Ruffini (2003e).

We can distinguish two different regimes corresponding respectively to $\gamma > 150$ and to $\gamma < 150$. In the E-APE region ($\gamma > 150$) the GRB substructure intensities indeed correlate with the ISM inhomogeneities. In this limited region (see peaks A, B, C) the Lorentz gamma factor of the ABM pulse ranges from $\gamma\sim 304$ to $\gamma\sim 200$. The boundary of the visible region is smaller than the thickness $\Delta R$ of the inhomogeneities (see Figs.~2,3 and Ruffini et al., 2002b, 2003b). Under these conditions the adopted spherical symmetry for the density spikes is not only mathematically simpler but also fully justified. The angular spreading is not strong enough to wipe out the signal from the inhomogeneity spike. The observational results reproduced in Fig.~3 present a remarkably improved fit, both in intensity, time scale and general morfology, with respect to the one considered in Ruffini et al. (2001c).

\begin{figure}
\plotone{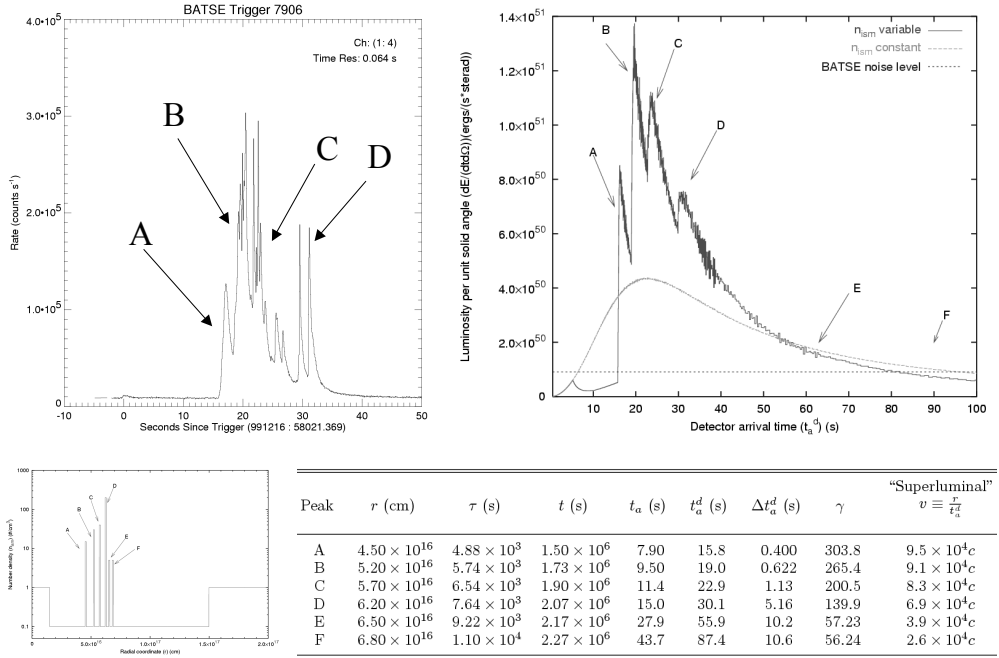}
\caption{Details of the fit of the E-APE structure of GRB~991216 in the EMBH theory. Details in Ruffini et al. (2002b, 2003b, 2003d)}
\end{figure}

Finally we then conclude:\\
{\bf a)} the informations carried by the afterglow, including the E-APE, are relevant for determining the structure of the ISM (Dermer's tomography);\\
{\bf b)} the main general relativistic effects are contained in the P-GRB. A new family of space missions are therefore urgently needed;\\
{\bf c)} for the first time we are witnessing the energy extraction from black holes.

\end{document}